\documentclass[12pt]{article}
\usepackage{graphicx}

\title{Seismic Full-Waveform Inversion Using Deep Learning Tools and Techniques}
\author{Alan Richardson (Ausar Geophysical)}
\begin{document}
\maketitle
\begin{abstract}
        I demonstrate that the conventional seismic full-waveform inversion algorithm can be constructed as a recurrent neural network and so implemented using deep learning software such as TensorFlow. Applying another deep learning concept, the Adam optimizer with minibatches of data, produces quicker convergence toward the true wave speed model on a 2D dataset than Stochastic Gradient Descent and than the L-BFGS-B optimizer with the cost function and gradient computed using the entire training dataset. I also show that the cost function gradient calculation using reverse-mode automatic differentiation is the same as that used in the adjoint state method.
\end{abstract}
\section{Introduction}
Deep learning is a method for training deep (many layer) neural networks (DNNs). It has recently attracted attention because of successful applications in several fields, including computer vision~\cite{lecun2015deep}. DNNs consist of multiple interconnected layers that each apply transformations to their inputs. When the output of a layer is fed back into the same layer as an input, it is said to be a recurrent neural network (RNN). Deep learning trains DNNs by inverting for the variables that parameterize the layer transformations. Reverse-mode automatic differentiation, including a backpropagation stage, is used to calculate the gradient of the chosen cost function with respect to the variables. The variables can then be iteratively updated so that the DNN produces the desired output.

Full-waveform inversion (FWI, \cite{virieux2017introduction}) is a method for deriving a model of subsurface properties such as wave speed from seismic data recorded on the surface. To achieve this, it performs nonlinear inversion iteratively, using the adjoint state method, including backpropagation, to determine the gradient \cite{plessix2006review}. The cost function is generally the sum of squares of the difference between the recorded data and data obtained by forward propagating a source wave through the current model.

There is clearly similarity between deep learning and FWI, but what is the benefit of combining them? Expressing an algorithm in the form of a DNN has the advantage of facilitating the use of deep learning software, such as TensorFlow~\cite{tensorflow2015-whitepaper}. These allow sophisticated implementations to be quickly developed. Once an algorithm is expressed using TensorFlow, for example, it can be executed on CPUs and GPUs, and run in parallel across a distributed memory computer cluster. Furthermore, it makes it easier to apply new ideas proposed in the active deep learning field to the algorithm, such as the Adam optimizer~\cite{kingma2014adam}.

My hypothesis is that the seismic algorithm FWI can be expressed using DNNs and so implemented using deep learning software, and that it may benefit from the deep learning practice of using minibatches of data with the Adam optimizer.

\section{Materials and Methods}
FWI requires a method for forwarding modeling -- numerically propagating a source wave through the model and recording the wavefield amplitude at receiver locations to create synthetic data. This allows the calculation of the cost function, the sum of the squared differences between this synthetic data and the true recorded data. FWI also requires a method for calculating the gradient of the cost function with respect to the wave speed model and any other model parameters used. This allows the model to be updated in a direction that reduces the cost function value, resulting, after multiple iterations, in a model that produces synthetic data that matches well with the true data.

I use TensorFlow for my implementation, but it should also be possible with other deep learning software.

In this section, I first describe the implementation of forward modeling of the wave equation using an RNN, and then discuss two approaches to calculating the gradient (automatic differentiation and the adjoint state method), followed by a description of the SEAM model that I use for experiments.

\subsection{Forwarding modeling}

Forward modeling can be implemented by repeatedly applying a sequence of arithmetic operations that correspond to one time step of a finite difference wave equation discretization. Sequences of arithmetic operations are directly representable using a neural network, where, in this case, the trainable variables are the subsurface model parameters such as wave speed. The network is therefore an RNN, and each repeating unit corresponds to one finite difference time step. For simplicity, I describe the 1D, constant density, acoustic (scalar wave equation) case. Extension to higher dimensions is straightforward. The implementation of more sophisticated wave equations is also, if numerical simulation of the chosen wave equation can be expressed using a sequence of operations permissible in neural networks.

In an RNN, the same operations are applied in each unit, but the data that these operations act upon changes. The data consist of state vectors, which are passed from one unit to the next and signify the current state of the system, and external inputs. In addition to passing state vectors to the next RNN unit, each unit also produces an output vector that is not passed on.

\begin{figure}
        \includegraphics{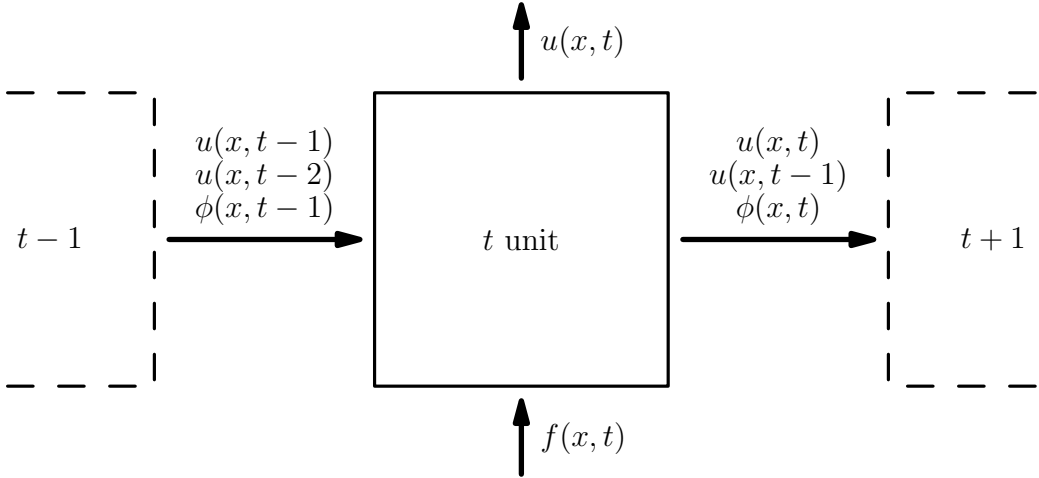}
        \caption{Wave equation forward modeling using an RNN. Each unit applies the operations to propagate forward one time step, taking the state from the previous unit (the wavefield $\mathbf{u}$ at adjacent time steps and auxiliary wavefield $\mathbf{\phi}$) and the source amplitude $\mathbf{f}$ as inputs, and producing updated state vectors (advanced by one time step) and the wavefield $\mathbf{u}$ as outputs.}
        \label{fig:rnn}
\end{figure}

When using a second-order finite difference approximation in time, two adjacent time steps of the wavefield need to be stored simultaneously. To prevent unwanted reflections from the edges of the simulation domain, I use a perfectly matched layer (PML). This requires the storage of an additional auxiliary wavefield. The state vector of the RNN therefore consists of two time steps of the regular wavefield, which I refer to as $u(x, t)$ and $u(x, t-1)$, and one auxiliary wavefield, which I call $\phi(x, t)$. The external input to each time step is the source amplitude at that time step, $f(x, t)$, and the output is either the full wavefield at time step $t$, $u(x, t)$, or only the wavefield amplitude at the receiver locations. This forward modeling neural network, using the conventional RNN structure, is depicted in Figure \ref{fig:rnn}.

In my implementation, I use convolution with a finite difference kernel to implement the spatial derivatives in the wave equation simulation. The convolution operation is optimized in DNN software due to its popularity in computer vision applications.

As wavefield values are only required at the receiver locations to compute the cost function, it is possible to only use those values as the output from the RNN units instead of the whole wavefield $u(x, t)$. In my implementation I output the whole wavefield as it is useful for the adjoint state method gradient calculation.

\subsection{Automatic differentiation}
As all of the operations needed to numerically simulate wave propagation are differentiable, automatic differentiation can be used to calculate the gradient of the cost function. It is thus only necessary to describe the forward modeling step of the algorithm and specify the cost function; the deep learning software then has sufficient information to calculate the gradient.

I derive the equations for automatic differentiation of the FWI cost function in Appendix~\ref{app:derivation}.

\subsection{Adjoint state method}
The traditional means of calculating the gradient in FWI is to use the adjoint state method. This involves backpropagating the difference between the forward modeled data and the recorded data, which can be achieved using the same forward modeling RNN described above. The equations of the adjoint state method and automatic differentiation are in fact the same, as I show in Appendix~\ref{app:derivation}. The difference is that I use the adjoint state method to refer to manually implementing the gradient computation rather than relying on the deep learning software's automatic differentiation.

Calculating the gradient of the cost function with respect to the wave speed is achieved using
\begin{equation}
        \frac{\partial J}{\partial \mathbf{c}} = \sum_{s=1}^{N_s} \sum_{t=1}^{N_t} -\frac{2}{\mathbf{c}^3}\frac{\partial^2 F\left(\mathbf{f}_{s, t}\right)}{\partial t^2}F^{-1}\left(2\mathbf{r}_{s, t}\right),
        \label{eqn:adjointstate}
\end{equation}
where $J$ is the cost function value, $\mathbf{c}$ is the wave speed model, $N_s$ and $N_t$ are the number of shots and time steps in the dataset, respectively, $\mathbf{f}_{s, t}$ contains the source amplitudes for shot $s$ as a function of time $t$, $\mathbf{r}_{s, t}$ is the difference between the synthetic and real receiver data for shot $s$ as a function of time $t$, and $F(\mathbf{g}_t)$ and $F^{-1}(\mathbf{g}_t)$ are wave propagators of the source amplitude $\mathbf{g}_t$ forward and backward in time, respectively. As $F\left(\mathbf{f}_{s, t}\right)$ is $u(x, t)$, the output from the forward modeling RNN, and $F^{-1}\left(2\mathbf{r}_{s, t}\right)$ uses the same forward modeling RNN but runs it backward in time and uses $2\mathbf{r}_{s, t}$ as the source term, we already have the necessary ingredients to implement the adjoint state method.

\subsection{SEAM dataset}
Some of the experiments that I describe in the results section use a model derived from the SEAM Phase I model~\cite{fehler2011seam}. I use a 2D section of the Vp model extracted from the 23900 m North line, covering from 7000 m to 16000 m in the horizontal direction and from 750 m to 6500 m in the depth direction. The extracted model is interpolated onto a grid with 100 m cell spacing. The dataset contains shots forward modeled on this model, using a 1 Hz Ricker wavelet as the source with a source and receiver spacing of one grid cell along the top surface.

In deep learning it is common to reserve part of the data as a ``development'' or ``validation'' dataset. This allows hyperparameters to be tuned with reduced risk of overfitting. In hyperparameter selection, models are trained using the ``training'' dataset (the remainder of the data, sometimes after also extracting a ``test'' dataset for final model testing) with various hyperparameter choices. The cost function values of the resulting trained models using the development dataset are used to select between the hyperparameter configurations, and, because this dataset was not used in training, we have greater confidence that this hyperparameter configuration will also result in a small cost function value on data not included in the training set; it generalizes well by not overfitting the data.

I reserve ten randomly selected shots for the development dataset, leaving the remaining eighty shots for training. The training shots are also shuffled so that minibatches of data are more likely to be representative of the whole dataset.

\section{Results}

To compare FWI implemented using deep learning concepts with a more conventional approach, I compare the gradients calculated using automatic differentiation and the adjoint state method, and perform inversion using the Adam and Stochastic Gradient Descent (SGD) optimizers (with minibatches) and the L-BFGS-B optimizer (using the entire dataset to calculate the cost and gradient), where the hyperparameters for the Adam and SGD optimizers were chosen using the development dataset.

The code to reproduce these results is contained in the ancillary archive.

\subsection{Gradient comparison}
The gradient of the cost function with respect to the model parameters is needed to update the model. In deep learning applications it is typically calculated by using automatic differentiation included in most deep learning software. In contrast, FWI applications generally use the adjoint state method to calculate the gradients. It is also possible (but computationally expensive) to use the finite difference method. This involves perturbing the model at each point and forward modeling using this perturbed model to determine the effect the perturbation has on the cost function. To compare these approaches, I create a random 1D model as the true model, randomly perturb it to produce an initial model, and compute the cost function gradient.

\begin{figure}
        \includegraphics{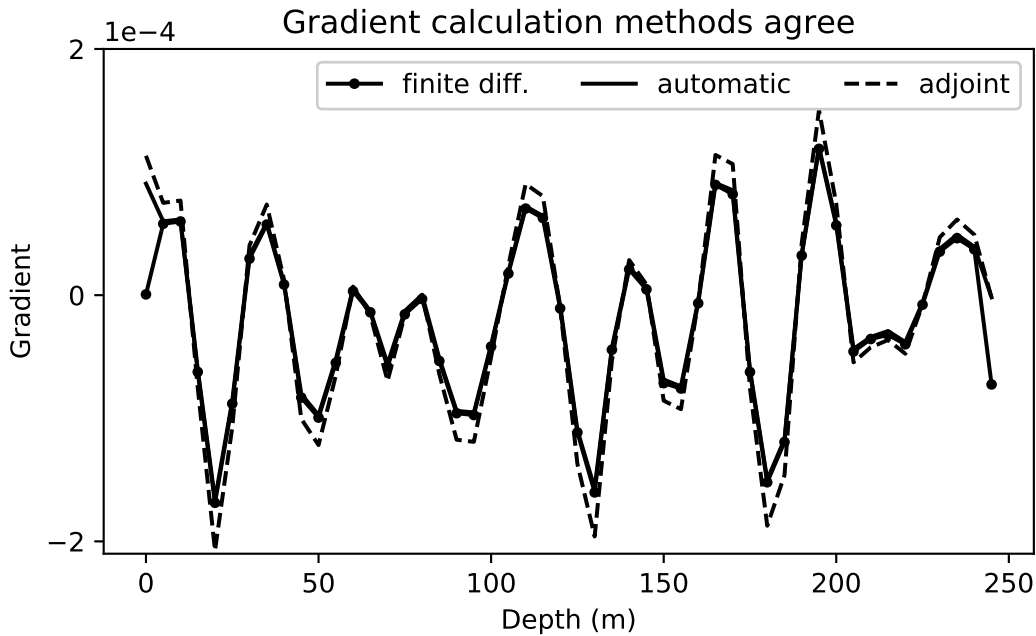}
        \caption{The gradient of the cost function with respect to the 1D wave speed model, computed using three approaches. The results are similar.}
        \label{fig:gradients}
\end{figure}

In Figure \ref{fig:gradients} I compare the gradient at each point in the 1D model computed using these three approaches. As expected, the results are similar. The adjoint state method does appear to slightly overestimate the gradients, and both it and automatic differentiation differ from the finite difference results at the edges of the model. These errors are likely to be due to numerical inaccuracies, and both are sufficiently small that they are unlikely to have a significant effect on iterative optimization.

\subsection{Using a development dataset for hyperparameter selection}

\begin{figure}
        \includegraphics{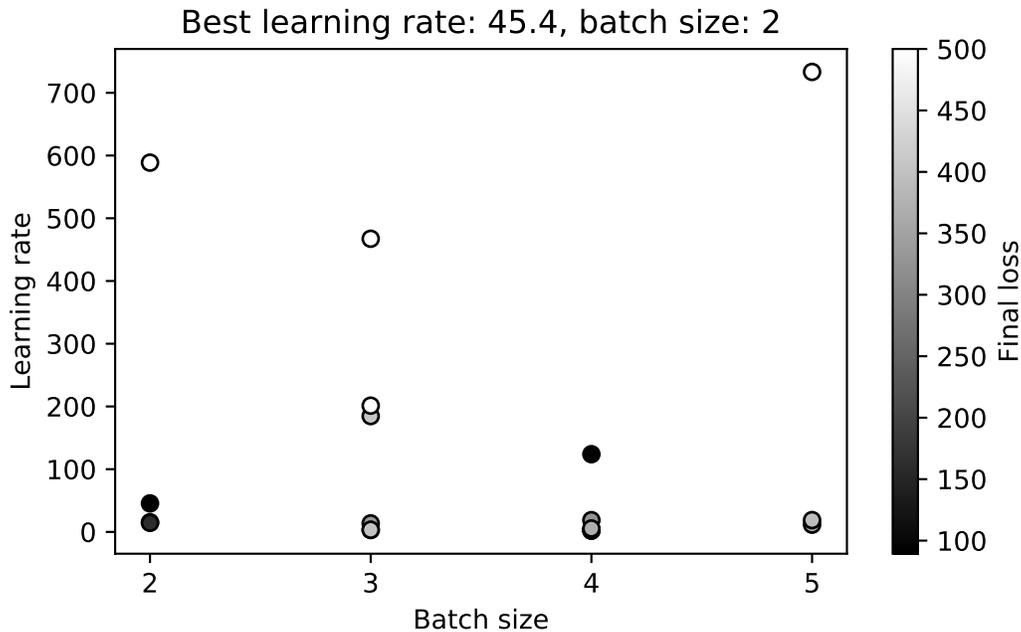}
        \caption{The cost function value (loss) on the development dataset after training using half of the training set with different combinations of batch size and learning rate hyperparameters using the Adam optimizer. Of the combinations considered, a learning rate of 45.4 and a batch size of 2 give the lowest loss.}
        \label{fig:hyperparam_selection}
\end{figure}

I wish to use minibatches of data with the Adam and SGD optimizers to perform training with my SEAM dataset. Two important hyperparameters are thus the batch size (the number of shots in each minibatch), and the learning rate (a number that controls the size of model updates). Randomly choosing twenty batch size and learning rate combinations, I train using each combination with approximately forty shots from the training dataset, and evaluate the cost function on the resulting models using the development dataset. The results for the Adam optimizer are plotted in Figure \ref{fig:hyperparam_selection}. It appears that for Adam, a learning rate between about 50 and 150 yields the lowest final cost function value. From these results, it is not possible to determine whether batch size has an effect. The lowest value results from a learning rate of 45.4 and a batch size of 2. The best combination for SGD is 733.35 and 5.

\subsection{Adam and SGD minibatch updates compared to L-BFGS-B}

Two particularly interesting techniques commonly used in deep learning are minibatches and the Adam optimizer~\cite{kingma2014adam}. Minibatches attempt to significantly reduce training time by updating the model after only computing the cost function and gradient on a small number of data samples. Computing the cost function and gradient using the entire dataset, a practice sometimes used in FWI, can ensure that each model update always reduces the cost function value, but this guarantee may not be worth the computational cost. The Adam optimizer uses concepts such as momentum to improve convergence and has been used successfully in many machine learning applications.

In this test I compare the Adam and SGD optimizers, using minibatches and the hyperparameters found in the previous section, with the L-BFGS-B optimizer~\cite{zhu1997algorithm}. For the latter, the cost and gradient are calculated using the entire dataset, and I constrain the range of inverted values to be between 1490 and 5000 m/s.

\begin{figure}
        \includegraphics{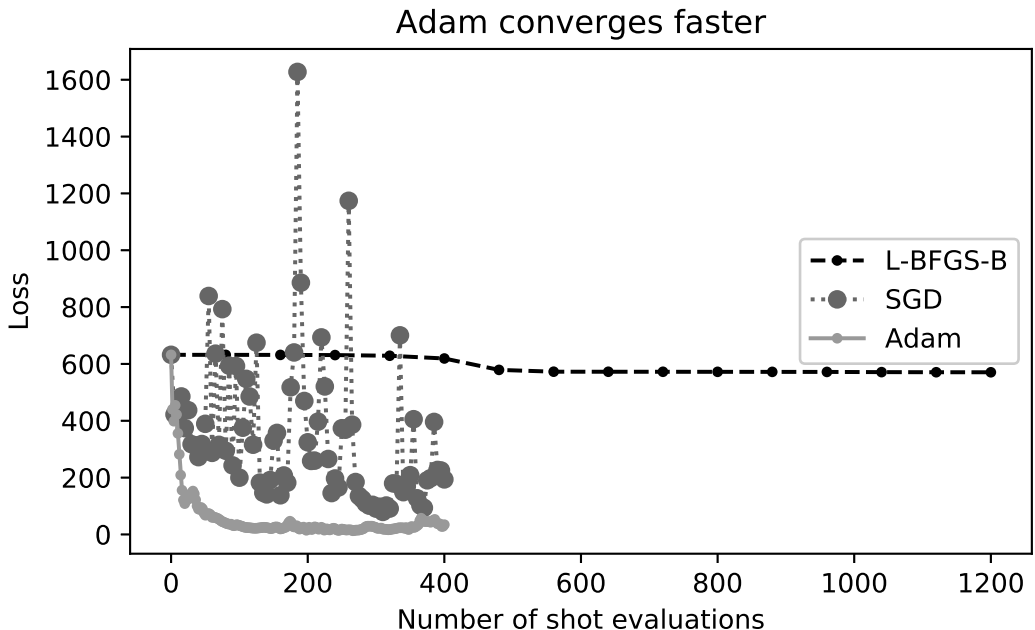}
        \caption{The cost function value (loss) on the development dataset declines more rapidly when using the Adam approach compared to SGD and L-BFGS-B. Adam is also more stable than SGD.}
        \label{fig:adam_vs_lbfgs_loss}
\end{figure}

Figure \ref{fig:adam_vs_lbfgs_loss} plots the cost function value calculated using the reserved development dataset on the current model of each optimizer as a function of the number of shot evaluations (calculations of the cost and gradient of one shot). It shows that Adam converges significantly more quickly than L-BFGS-B. Adam's convergence is also quicker and more stable than SGD. In fact, the cost function when using Adam becomes flat after approximately one pass through the training dataset. This means that even when the additional shot evaluations used to determine the hyperparameters for Adam are considered, it still converges more quickly than L-BFGS-B did in this experiment.

\begin{figure}
        \includegraphics{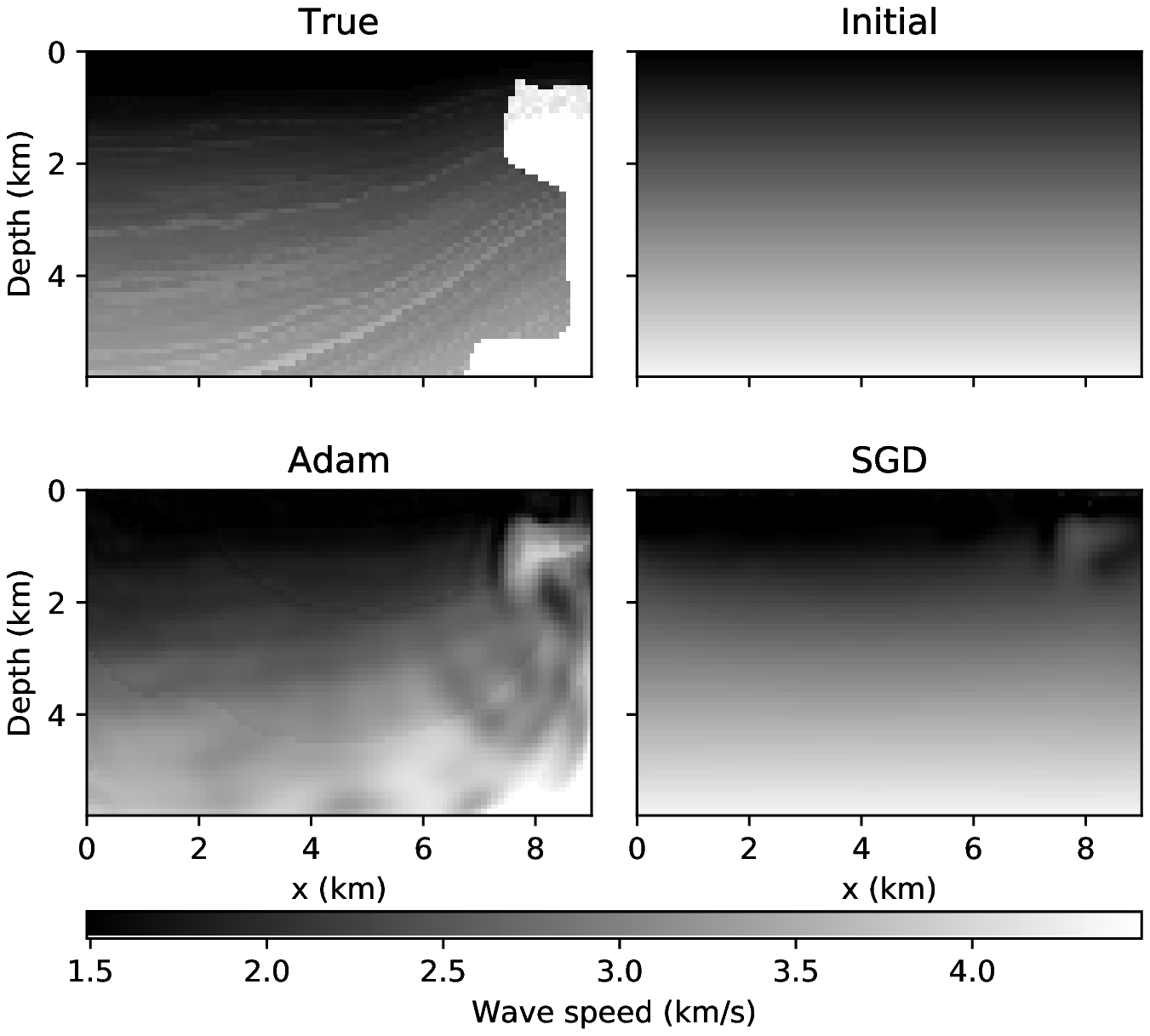}
        \caption{The model produced by Adam is close to the true model, while SGD has only captured shallow features.}
        \label{fig:adam_vs_lbfgs_models}
\end{figure}

The final models produced by Adam and SGD after 400 shot evaluations (not including those used for hyperparameter selection), are shown in Figure \ref{fig:adam_vs_lbfgs_models}. As expected, given its low final cost function value, the model produced by Adam is quite similar to the true model: the sediment areas are mostly well matched, and parts of the high velocity salt body on the right of the model are approximately correct. Some shallow features, such as the top of the salt body and the sea floor, are visible in the SGD model, but deeper parts of the model are incorrect. The model produced by L-BFGS-B after 1200 shot evaluations is not shown as it has not changed significantly from the initial model. This is not surprising, as it has only evaluated 16 different models, while the Adam model was produced after evaluating 201 different models (not including the hyperparameter selection process).

\section{Discussion}
\paragraph{Automatic differentiation} There are several advantages of using automatic differentiation to calculate the gradient of the cost function with respect to model parameters. Foremost among these is that it greatly reduces development time as it avoids the need to manually implement and verify the adjoint state method. Another advantage is that it makes it easier to try new ideas, such as experimenting with other deep learning concepts like dropout~\cite{srivastava2014dropout}. The major disadvantage, however, is its high memory requirement, as it stores all of the wavefield time slices in memory. In FWI, the wavefield at the final time step can certainly be affected by interactions of the wavefield and the model during the initial time steps, and so backpropagation must occur over the entire sequence of time steps. It is not typical in deep learning applications to backpropagate through thousands of layers. As a result, automatic differentiation in deep learning software is not usually designed to efficiently handle such situations, and does not employ strategies common in FWI software to reduce memory requirements. These strategies include not saving the wavefield at every time step, applying lossy compression to the wavefield time slices that are stored, saving the wavefields to disk instead of keeping them in memory, and regenerating wavefields during backpropagation rather than storing them. Using the adjoint state method to compute the gradient instead of relying on automatic differentiation provides enough control, even when using deep learning software, to implement some of these strategies. This means that the adjoint state method, rather than automatic differentiation, is likely to be necessary when applying FWI to realistically sized models.

\paragraph{Adam} Instead of waiting to check every shot before updating the model, minibatch updates allow more rapid exploration of the model space. One of the main advantages of the Adam optimizer is its suitability for working with minibatches. Although it is possible to use minibatches with some more traditional optimizers, as in Stochastic Gradient Descent~\cite{diaz2011fast}, Adam's use of concepts such as momentum make it particularly well adapted to minibatches. Momentum not only helps to improve convergence by reducing the propensity of optimizers to struggle in narrow valleys of the cost function, but it also tackles the problem of small minibatches not being representative of the entire dataset by smoothing gradients over several minibatches. This is likely to be the cause of Adam's greater stability compared to SGD in the SEAM experiment. A further attraction of Adam is its wide availability in deep learning software. In contrast, TensorFlow does not provide an L-BFGS optimizer, and so I had to use L-BFGS-B from another software package, SciPy~\cite{scipy}. An advantage of L-BFGS algorithms is that they have few or no hyperparameters. Choosing hyperparameters for Adam and SGD required running many trial inversions. In this example, even when these trial inversions are included, Adam converges more quickly than L-BFGS-B, but this may not always be the case. A recently proposed modification of L-BFGS allows it to also be used with minibatches \cite{fabien2017stochastic}, which may make it more competitive with Adam.

\paragraph{Development datasets} Excluding part of the data from training for use as a development dataset is standard practice in deep learning, but is likely to face resistance in the FWI community. Part of the reason for the opposition is probably due to seismic datasets generally having fewer data samples than those used in deep learning applications, relative to the number of model parameters. The example of selecting the Adam and SGD hyperparameters demonstrates that the practice can result in fast convergence to a good model, but does not prove that a similar result could not have been achieved using the entire dataset. The necessity of a development dataset is more obvious when choosing regularization-related hyperparameters, as in that situation the risk of overfitting is clearer.

\paragraph{Other applications of deep learning to FWI} The deep learning approach to FWI I describe here is effectively the conventional FWI algorithm implemented using deep learning software and using several concepts popular in deep learning. There are other ways of applying deep learning techniques to seismic inversion that have less resemblance to traditional FWI. One of these is to use an approach more obviously inspired by computer vision applications, where the DNN is trained to transform seismic data into a model. With this approach, inversion only occurs during training. Once the network has been trained, it can be applied to datasets without performing wave propagation. Such a technique would be particularly useful for specific tasks such as salt body reconstruction \cite{lewis2017deep} and obtaining a low-wavenumber starting model for traditional FWI \cite{araya2018deep}.

\section{Conclusion}
The hypothesis that FWI can be implemented using an RNN with deep learning software has been verified. Automatic differentiation calculates gradients similar to those produced by the traditional adjoint state method approach, but as the implementation of it in deep learning software is typically not designed for RNNs as deep as those needed for FWI, the memory requirements are likely to be prohibitive for realistic applications. The hyperparameters that were found using a separate development dataset appear to have been good choices for inverting the example dataset. Combined with the Adam optimizer and the concept of minibatches, these resulted in significantly faster convergence than when using the L-BFGS-B optimizer with the cost and gradient calculated using the entire dataset, and even performed better than SGD.

\section{Acknowledgments}
I am grateful to Bram Willemsen for reviewing an earlier version of this article.

\bibliography{seismic_dnn}{}
\bibliographystyle{plain}
\appendix
\section{Derivation of automatic differentiation procedure}
\label{app:derivation}
In this appendix I derive the automatic differentiation procedure for calculating the gradient of the cost function with respect to the model parameters. For simplicity, I use the 1D scalar wave equation, ignore the non-reflecting PML regions, work in discretized space and time coordinates, and assume that the data consist of only one shot with one receiver.

Using sum of squared errors as the cost function,
\begin{equation}
        J = \sum_{t=1}^{N_t} (\mathbf{\delta_{x_r}}^T\mathbf{u}_t - d_t)^2,
\end{equation}
where $J$ is the scalar cost function value, $N_t$ is the number of time steps, $\mathbf{u}_t$ is a column vector containing the forward propagated wavefield at time step $t$, $\mathbf{\delta_{x_r}}$ is a column vector that is only non-zero at the location of the receiver (its purpose is to extract the wavefield amplitude at the receiver), and $d_t$ is a scalar containing the true receiver amplitude at time step $t$.

The cost function depends on the wave speed model implicitly through $\mathbf{u}_t$. When using a second-order accurate central finite difference scheme to approximate the second-order time derivative in the wave equation, the wavefield depends on the wave speed, $\mathbf{c}$, and the two previous time steps, $\mathbf{u}_{t-1}$, and $\mathbf{u}_{t-2}$,
\begin{equation}
        \mathbf{u}_t = \mathbf{c}^2 \Delta_t^2\left(\mathbf{D_x}^2\mathbf{u}_{t-1} - \mathbf{f}_{t-1}\right) + 2\mathbf{u}_{t-1} - \mathbf{u}_{t-2},
        \label{eqn:waveeqn}
\end{equation}
where $\Delta_t$ is the time step size, $\mathbf{D_x}^2$ is a symmetric, square matrix that applies the second spatial derivative, and $\mathbf{f}_{t-1}$ is a column vector containing the shot's source amplitude at time step $t-1$ as a function of space.

Reverse-mode automatic differentiation uses the chain rule to calculate the derivative of the cost function value with respect to variables as it proceeds backward through the computation graph. Because the intermediate variables, $\mathbf{u}_t$, have dependencies between themselves, it is important to note which variables are being held constant in the partial derivative notation. The gradient of the cost function with respect to one wavefield time step is
\begin{eqnarray}
        \nonumber\left.\frac{\partial J}{\partial \mathbf{u}_t}\right|_{\mathbf{c}} &=&
        \left.\frac{\partial J}{\partial \mathbf{u}_t}\right|_{\mathbf{c}, \mathbf{u_{t' \neq t}}}
        + \left.\frac{\partial J}{\partial \mathbf{u}_{t+1}}\right|_{\mathbf{c}}\left.\frac{\partial \mathbf{u}_{t+1}}{\partial \mathbf{u}_t}\right|_{\mathbf{c}, \mathbf{u_{t' \neq t, t+1}}}\\
        &&+ \left.\frac{\partial J}{\partial \mathbf{u}_{t+2}}\right|_{\mathbf{c}}\left.\frac{\partial \mathbf{u}_{t+2}}{\partial \mathbf{u}_t}\right|_{\mathbf{c}, \mathbf{u_{t' \neq t, t+2}}},
\end{eqnarray}
where $\left.\frac{\partial J}{\partial \mathbf{u}_t}\right|_{\mathbf{c}, \mathbf{u_{t' \neq t}}}$ indicates the row vector of partial derivatives of $J$ with respect to elements of $\mathbf{u}_t$, while the wave speed, $\mathbf{c}$, and the wavefields from other time steps, $\mathbf{u_{t' \neq t}}$, are held constant. Variables are only held constant for the partial derivative noted. The partial derivative vector $\left.\frac{\partial J}{\partial \mathbf{u}_{t+1}}\right|_{\mathbf{c}}$, for example, includes all changes to $J$ caused by changes in $\mathbf{u}_{t+1}$, including those due to dependencies of later wavefield time steps on $\mathbf{u}_{t+1}$.

In a similar way, automatic differentiation calculates the gradient of the cost function with respect to the wave speed by accumulating terms of the sum that can be evaluated using the chain rule as the method progresses backward through time steps,
\begin{equation}
        \frac{\partial J}{\partial \mathbf{c}} = \sum_{t=1}^{N_t} \left.\frac{\partial J}{\partial \mathbf{u}_t}\right|_\mathbf{c} \left.\frac{\partial\mathbf{u}_t}{\partial \mathbf{c}}\right|_{\mathbf{u_{t' \neq t}}}.
        \label{eqn:accum_djdc}
\end{equation}

As an example, if $N_t=4$, the steps to calculate the gradient of $J$ would be
\begin{equation}
        \left.\frac{\partial J}{\partial \mathbf{u}_4}\right|_{\mathbf{c}} =
        \left.\frac{\partial J}{\partial \mathbf{u}_4}\right|_{\mathbf{c}, \mathbf{u_{t' \neq 4}}},
\end{equation}\begin{equation}
        \left.\frac{\partial J}{\partial \mathbf{u}_3}\right|_{\mathbf{c}} =
        \left.\frac{\partial J}{\partial \mathbf{u}_3}\right|_{\mathbf{c}, \mathbf{u_{t' \neq 3}}}
        + \left.\frac{\partial J}{\partial \mathbf{u}_{4}}\right|_{\mathbf{c}}\left.\frac{\partial \mathbf{u}_{4}}{\partial \mathbf{u}_3}\right|_{\mathbf{c}, \mathbf{u_{t' \neq 3, 4}}},
\end{equation}\begin{equation}
        \left.\frac{\partial J}{\partial \mathbf{u}_2}\right|_{\mathbf{c}} =
        \left.\frac{\partial J}{\partial \mathbf{u}_2}\right|_{\mathbf{c}, \mathbf{u_{t' \neq 2}}}
        + \left.\frac{\partial J}{\partial \mathbf{u}_{3}}\right|_{\mathbf{c}}\left.\frac{\partial \mathbf{u}_{3}}{\partial \mathbf{u}_2}\right|_{\mathbf{c}, \mathbf{u_{t' \neq 2, 3}}}
        + \left.\frac{\partial J}{\partial \mathbf{u}_{4}}\right|_{\mathbf{c}}\left.\frac{\partial \mathbf{u}_{4}}{\partial \mathbf{u}_2}\right|_{\mathbf{c}, \mathbf{u_{t' \neq 2, 4}}},
\end{equation}\begin{equation}
        \left.\frac{\partial J}{\partial \mathbf{u}_1}\right|_{\mathbf{c}} =
        \left.\frac{\partial J}{\partial \mathbf{u}_1}\right|_{\mathbf{c}, \mathbf{u_{t' \neq 1}}}
        + \left.\frac{\partial J}{\partial \mathbf{u}_{2}}\right|_{\mathbf{c}}\left.\frac{\partial \mathbf{u}_{2}}{\partial \mathbf{u}_1}\right|_{\mathbf{c}, \mathbf{u_{t' \neq 1, 2}}}
        + \left.\frac{\partial J}{\partial \mathbf{u}_{3}}\right|_{\mathbf{c}}\left.\frac{\partial \mathbf{u}_{3}}{\partial \mathbf{u}_1}\right|_{\mathbf{c}, \mathbf{u_{t' \neq 1, 3}}},
\end{equation}
where after each step we calculate the term of Equation~\ref{eqn:accum_djdc} corresponding to the current time step.

Calculating the required partial derivatives, we obtain
\begin{equation}
        \left.\frac{\partial J}{\partial \mathbf{u}_t}\right|_{\mathbf{c}, \mathbf{u_{t' \neq t}}} = 2(\mathbf{\delta_{x_r}}^T\mathbf{u}_t - d_t)\mathbf{\delta_{x_r}}^T,
\end{equation}
\begin{equation}
        \left.\frac{\partial \mathbf{u}_{t+1}}{\partial \mathbf{u}_t}\right|_{\mathbf{c}, \mathbf{u_{t' \neq t, t+1}}} = \mathbf{c}^2 \Delta_t^2 \mathbf{D_x}^2 + 2,
\end{equation}
\begin{equation}
\left.\frac{\partial \mathbf{u}_{t+2}}{\partial \mathbf{u}_t}\right|_{\mathbf{c}, \mathbf{u_{t' \neq t, t+2}}} = -1,
\end{equation}
\begin{equation}
        \left.\frac{\partial\mathbf{u}_t}{\partial \mathbf{c}}\right|_{\mathbf{u_{t' \neq t}}} = 2 \mathbf{c} \Delta_t^2 \left( \mathbf{D_x}^2 \mathbf{u}_{t-1} - \mathbf{f}_{t-1}\right).
\end{equation}

Substituting into Equation \ref{eqn:accum_djdc}, we find
\begin{eqnarray}
        \nonumber\frac{\partial J}{\partial \mathbf{c}} &=& \sum_{t=1}^{N_t} \left(2(\mathbf{\delta_{x_r}}^T\mathbf{u}_t - d_t)\mathbf{\delta_{x_r}}^T
        + \left.\frac{\partial J}{\partial \mathbf{u}_{t+1}}\right|_{\mathbf{c}}(\mathbf{c}^2 \Delta_t^2 \mathbf{D_x}^2 + 2)
        - \left.\frac{\partial J}{\partial \mathbf{u}_{t+2}}\right|_{\mathbf{c}}\right) \\
        &&\times 2 \mathbf{c} \Delta_t^2 \left( \mathbf{D_x}^2 \mathbf{u}_{t-1} - \mathbf{f}_{t-1}\right)\\
        \nonumber &=& \sum_{t=1}^{N_t} \left(\mathbf{c}^2 \Delta_t^2 \left(\mathbf{D_x}^2\left.\frac{\partial J}{\partial \mathbf{u}_{t+1}}\right|_{\mathbf{c}} + \frac{2}{\mathbf{c}^2 \Delta_t^2} (\mathbf{\delta_{x_r}}^T\mathbf{u}_t - d_t)\mathbf{\delta_{x_r}}^T\right)\right.\\
        && + 2\left.\left.\frac{\partial J}{\partial \mathbf{u}_{t+1}}\right|_{\mathbf{c}}
        - \left.\frac{\partial J}{\partial \mathbf{u}_{t+2}}\right|_{\mathbf{c}}\right)
        2 \mathbf{c} \Delta_t^2 \left( \mathbf{D_x}^2 \mathbf{u}_{t-1} - \mathbf{f}_{t-1}\right)\\
        &=& \sum_{t=1}^{N_t} -F^{-1}\left(\frac{2}{\mathbf{c}^2 \Delta_t^2} (\mathbf{\delta_{x_r}}^T\mathbf{u}_t - d_t)\mathbf{\delta_{x_r}}^T\right)
        2 \mathbf{c} \Delta_t^2 \left( \mathbf{D_x}^2 \mathbf{u}_{t-1} - \mathbf{f}_{t-1}\right),
\end{eqnarray}
where $F^{-1}(\mathbf{g}_t)$ is wave propagation backward in time of the source amplitude $\mathbf{g}_t$. We also recognize that the wave equation can be used to express the factor on the right as the second time derivative of the forward propagated source wavefield (see Equation \ref{eqn:waveeqn}),
\begin{eqnarray}
        \frac{\partial J}{\partial \mathbf{c}} &=&\sum_{t=1}^{N_t}-F^{-1}\left(\frac{2}{\mathbf{c}^2 \Delta_t^2} (\mathbf{\delta_{x_r}}^T\mathbf{u}_t - d_t)\mathbf{\delta_{x_r}}^T\right)\frac{2\Delta_t^2}{\mathbf{c}}\frac{\partial^2\mathbf{u}_{t-1}}{\partial t^2}\\
        &=&\sum_{t=1}^{N_t}-F^{-1}\left(2 (\mathbf{\delta_{x_r}}^T\mathbf{u}_t - d_t)\mathbf{\delta_{x_r}}^T\right)\frac{2}{\mathbf{c}^3}\frac{\partial^2\mathbf{u}_{t-1}}{\partial t^2},
\end{eqnarray}
where I used the linearity of wave propagation to take factors out of the source amplitude of the left term. The result is the same equation that is used in the adjoint state method (Equation \ref{eqn:adjointstate}).
\end{document}